# Modeling the guaranteed delivery of bulk data

Mauro Femminella and Gianluca Reali
DIEI, University of Perugia, Perugia, Italy
{mauro.femminella, gianluca.reali}@diei.unipg.it

Roberto Francescangeli
CS – Columbia University, New York, USA
roberto.francescangeli@gmail.com

*Abstract*—**The delivery of bulk data is an increasingly pressing problem in modern networks. While in some cases these transfers happen in background without specific constraints in terms of delivery times, there are a number of scenarios in which the transfer of tens of GB of data must occur in specific, limited time windows. In order to face this task, a suitable solution is the deployment of virtual links with guaranteed bandwidth between endpoints provided by a Service Overlay Network (SON) provider. We model this scenario as an optimization problem, in which the target consists of minimizing the costs of the virtual links provided by the SON and the unknowns are the provisioned bandwidths of these links. Since the resulting objective function is neither continuous nor convex, the solution of this problem is really challenging for standard optimization tools in terms of both convergence time and solution optimality. We propose a solution based on an heuristic approach which uses the min-plus algebra. Numerical results show that the proposed heuristic outperforms the considered optimization tools, whilst maintaining an affordable computation time.**

## I. INTRODUCTION

The delivery of bulk data is an increasingly pressing problem in different network services, such as those enabling content distribution networks (CDN) [1], (mobile) cloud computing [29][38], Digital Cinema distribution [2], and so on. Although in some cases file transfer may happen in background without specific constraints in terms of delivery times (e.g. distribution of data backups), in some other challenging scenarios, transfer of (tens of) GBs of data must occur in a limited timeframe, which could be in the order of minutes or hours. Hence, these kind of network service can be referred to as guaranteed delivery of bulk data. Typical use cases include content update in CDN edge servers, for instance for pre-loading new HD videos when a flash crowd of requests is expected, migration/deployment of virtual machines in remote datacenters [38], distribution of digital cinema files to theatres the night before projection.

We face this problem from the viewpoint of a generic *content operator*. We assume that the main goal to achieve is having the desired contents correctly stored and available to customers in all sites at a given time. In order to solve this problem, content operators make typically use of a number of tunnels with guaranteed bandwidth connecting content sources and content destinations. In this way data transfer time is predictable. Such tunnels constitute a so-called *hybrid virtual private network* (VPN) in the terminology defined by the VPN Consortium [9] or a *provider-provisioned VPN* (PP-VPN) in that of RFC 4110 [11]. Deployment of hybrid VPNs with guaranteed bandwidth is clearly not an activity typical of a content operator. Thus, we assume the presence of a Service Overlay Network (SON) provider, that runs the business activity of providing content operators with hybrid VPNs on-demand. Thus, the ultimate objective is to minimize the cost of these tunnels, having the desired contents stored in all target locations at a given time. In turn, a SON needs to buy resources from underlying Internet Service Providers (ISPs) so as to provide his customers with virtual paths with guaranteed bandwidths. Thus, we need to consider a three layer architecture, illustrated in Fig. 1. Since our contribution is part of the distribution layer, we focus only on it and do not deal with details relevant to lower layers, already widely treated in literature, such as internal structure and algorithms of overlay networks [10][21][22][24]. Also, both suitability of existing transfer protocols to support efficient transfer of bulk data ([12][13][14]) and solutions to secure SON tunnels ([15][23]) are topics which have been widely addressed in the literature and are considered out the scope of this paper. We assume that for each pair of endpoints <content source, content destination>, a SON provider is able to provide information about the amount of bandwidth that could be guaranteed for connecting them (i.e., the maximum VPN capacity).

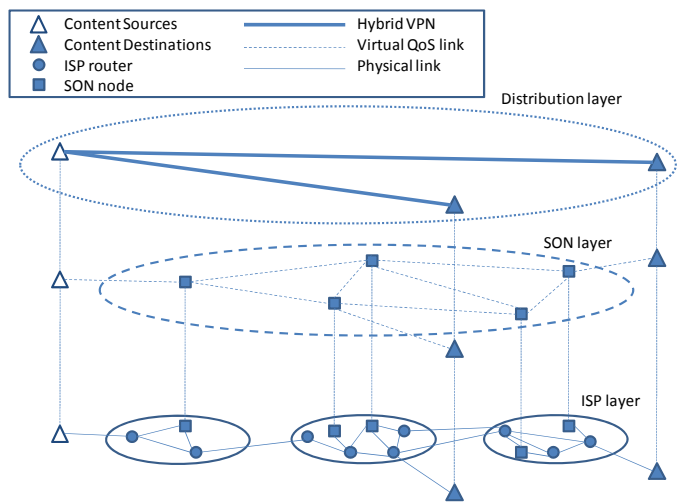

Fig. 1. Distribution architecture at different layers.

We model this scenario through an optimization problem, in which the target consists of minimizing the costs of the virtual paths provided by the SON and the unknowns are the provisioned bandwidths of these paths. We assume, without restriction of generality, that the content to distribute is present in either one or more sources. This is quite common in many scenarios, such as the simultaneous availability of the same virtual machines in multiple datacenters for disaster recovery, as illustrated in [28]. This feature allows implementing parallel downloading techniques [5], giving to the content operator larger degrees of freedom and solution space.

Under the realistic assumptions, we show that the objective function may be neither continuous nor convex. Under these conditions, the solution of this problem is really challenging for standard optimization tools in terms of both convergence time and solution optimality. To overcome these issues, we propose a solution based on an heuristic approach which uses the min-plus algebra. Numerical results show that the proposed heuristic outperforms the considered optimization tools, whilst maintaining an affordable computation time.

The paper is organized as follows. In Section II, we illustrate the mathematical model of the above distribution system. In Section III we illustrate the proposed heuristic solution based on the min-plus algebra. In section IV we show a comparison between the results obtained with our heuristic and the those obtained with standard optimization solver tools. In section V, we discuss our finding and relate them to other works in the field. Finally, in Section VI, we draw our concluding remarks.

## II. BANDWIDTH ALLOCATION PROBLEM FORMULATION

In this Section, we illustrate the mathematical model of the system. Below we report inputs, outputs, constraints and optimization function of the optimization problem.

### A. Inputs to the problem

- $K$: Number of content source (indexed by $j$).
- $N$: Number of content destinations (indexed by $i$).
- $H$: Number of data item (indexed by $l$) in the system.
- $D_l$: Size of the data item $l$.
- $C_i^T$: Current available bandwidth on the access link of destination $i$.
- $C_j^M$: Current available bandwidth on the access link of content source $j$.
- $\bar{c}_{ji}$: Maximum bandwidth which can be allocated for the VPN from source $j$ to destination $i$.
- $\tau_{il}$: Maximum download for film $l$ required by destination $i$ ($\tau_{il} \geq D_l / C_i^T$, $\forall(i,l)$)
- $u_{il}$: Request indicator, it is equal to 1, if destination $i$ has requested data item $l$, otherwise it is set to 0
- $v_{jl}$: Presence indicator, it is equal to 1 if source $j$ stores data item $l$, otherwise it is set to 0

### B. Outputs of the problem:

- $c_{jil}$: Capacity allocated from source $j$ to destination $i$ to download (eventually a part of) data item $l$
- $D_{jil}$: Size of the fragment of data item $l$ to be retrieved by destination $i$ from source $j$
- $c_{ji}$: $c_{ji} = \sum_{l=1}^{H} c_{jil}$ is the capacity allocated to the VPN from source $j$ to destination $i$

### C. Problem constraints

The problem is characterized by the following constraints:

$$\sum_{i=1}^{N} c_{ji} \leq C_j^M, \quad j=1,..,K ; \quad (1)$$

$$\sum_{j=1}^{K} c_{ji} \leq C_i^T, \quad i=1,..,N ; \quad (2)$$

$$\sum_{j=1}^{K} D_{jil} = D_l u_{il}, \quad \forall(i,l) ; \quad (3)$$

$$D_{jil} \leq \tau_{il} \cdot c_{jil}, \quad \forall (j,i,l) ; \quad (4)$$

$$c_{ji} \leq \bar{c}_{ji}, \quad \forall(j,i) ; \quad (5)$$

$$0 \leq c_{jil} \leq v_{jl} \cdot u_{il} \cdot \bar{c}_{ji}, \quad \forall(j,i,l) ; \quad (6)$$

$$0 \leq D_{jil} \leq v_{jl} \cdot u_{il} \cdot D_l, \quad \forall(j,i,l) \quad (7)$$

Equations (1) and (2) represent the bandwidth constraint on source and destination access links, respectively. The integrity constraint on each requested data item is given by (3), whereas the maximum download time constraint for each data item request is set by means of (4). The bandwidth constraint on each VPN ($j \rightarrow i$) is represented by (5). Finally, constraints (6) and (7) enforce a value equal to 0 on reserved bandwidth ($c_{jil}$) and fragment size ($D_{jil}$) for data item $l$ between source $j$ and destination $i$, if the relevant data item $l$ is not requested by destination $i$ or is not available in source $j$.

We recall that the amount of allocated bandwidth on each VPN is

$$c_{ji} = \sum_{l=1}^{H} c_{jil} . \quad (8)$$

Please note that $c_{ji}$ is the amount of bandwith to buy from the SON and thus it is the real output of the system, whereas $c_{jil}$ is the bandwith contribution to $c_{ji}$ given by the download of data item $l$ on that VPN.

### D. Objective function of the problem

The objective cost function to be minimised is

$$F_{TOT} = \sum_{(j,i)} f_{ji}(c_{ji}), \quad (9)$$

where $f_{ji}(c_{ji})$ is the cost of the VPN $(j,i)$. This function has to be non-decreasing with the amount of reserved capacity $c_{ji}$, and obviously $f_{ji}(0)=0$. In general, it depends on the specific $(j,i)$ pair and can be written as follows:

$$f_{ji}(c_{ji}) = \begin{cases} a_{ji} + g_{ji}(c_{ji}) & c_{ji} > 0 \\ 0 & c_{ji} = 0 \end{cases}, \quad \forall (j,i) \quad (10)$$

where the parameter $a_{ji}>0$ accounts for the VPN $(j,i)$ set-up and maintenance cost[1] and $g_{ji}(c_{ji})$ is the cost contribution accounting for reserved resources on the VPN. The $g_{ji}()$ function has to be positive, sub-additive, and non-decreasing with $c_{ji}$, and $g_{ji}(0)=0$. We choose $g_{ji}(c_{ji})$ to be sub-additive, since, due to economies of scale, the operational/commercial cost of network resources is typically sub-additive. For the same reason, the authors in [21] suggest to use *concave* cost functions with respect to provisioned bandwidth. Thus, $f_{ji}$ is non-convex, discontinuous at $c_{ji}$=0 (with $f_{ji}(0)$=0), non-decreasing, and sub-additive.

---

[1] Each edge device of the SON managing provisioned VPNs must maintain a separate, logical body (Virtual Forwarding Instance, VFI) for each connected VPN [11]. A VFI contains a router information base and a forwarding information base for each VPN. Thus, an edge device uses an amount of network and computing resources in proportion to the number of VFIs. This limits the number of VPNs which an edge device can support, and thus it is reasonable that just the set-up of a VPN has a cost, independently of the provisioned resources.

Note that all the outputs reported above can be computed by combining the $c_{jil}$ values; clearly, if $c_{jil}$>0 the source $j$ has been selected to satisfy (a part of) the request $u_{il}$. In fact, $c_{jl}$ can be expressed as a combination of $c_{jil}$. In addition, $D_{jil}$ can be computed as $\tau_{il} c_{jil}$ if the download time is reasonably set to the maximum achievable value $\tau_{il}$, minimising the bandwidth consumption. This occurs when $f_{ji}(c_{ji})$ strictly increases with $c_{ji}$ and/or the constraint (4) is set as an equality. Thus, in order to solve the system, we have to calculate $c_{jil}$.

The obtained cost function (9) is thus a linear combination of discontinuous, non-convex functions, and thus the problem is neither convex nor continuous. This suggests that convergence to the global optimum cannot be guaranteed, and that the usage of heuristics methods is justified.

### III. THE MPH HEURISTIC

The intuition underlying the proposed formulation is that, *for the case of the single request, it provides an elegant and effective way to define an exhaustive search method within the set of admissible solutions*. If there are more requests, different ways to solve the problem exist. The first is using classic optimization techniques, which however cannot be successfully applied to this problem, since the objective function is not convex, non linear, and not continuous. A second option could be to use an exhaustive search method to the problem in equation (9) as a whole. This way is mathematically intractable, an example will be provided in section IV. The third one, which we adopt, consists of (i) selecting the service requests one at time, (ii) calculating the transmission resources needed at each node to provide the content within the desired download time, (iii) update the available transmission resources, and go on with another service request until they are exhausted. Thus, it is necessary to design an algorithm which selects the service request at each computation cycle so as to minimize the cost function.

In more detail, we propose a greedy, heuristic algorithm (Min-Plus Heuristic, MPH). Optimal solutions are searched by applying rules defined by means of min-plus algebra [6]. The flow diagram of the algorithm is depicted in Fig. 2. The key point is the algorithm that selects the request for which calculating the VPNs to set up at each computation cycle. As shown in Fig. 2, our solution uses a centralized decision system, which collects service requests, network status, and finally takes decisions about which on-demand VPNs to set-up. The inputs of the algorithm are the service requests collected at a given time, the network status (bandwidth availability) and the data item catalogue available per-source, and the output are the VPN endpoints and their bandwidth.

The algorithm is organised in *cycles*, and for each cycle one data item request, $u_{il}$, is served. This means that for each cycle the set of sources from which the data item will be downloaded has to be determined and the bandwidth has to be allocated to the relevant VPNs according to the required download time. Clearly, the network status is updated at the beginning of each cycle by considering the amount of resources allocated in the previous one. The algorithm ends when all requests are served.

In this Section, we first describe the single-request routing algorithm based on the min-plus convolution, and then we present a set of possible scheduling criteria.

In the next Section, we will compare the performance of MPH with that of two commercial solvers, LINGO [7], and MINOS [8], a solver for non-linear optimization problems using the AMPL modelling language, in both homogeneous (all service request are equal, i.e. the same download time, the same data item requested or the requested data items are of the same size and are present in the same source nodes) and heterogeneous scenarios.

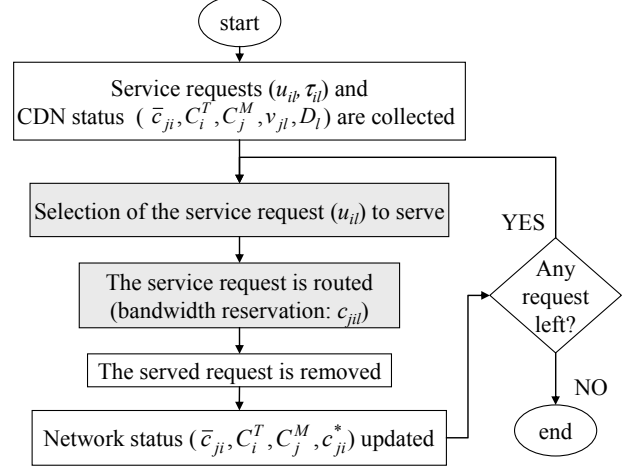

Fig. 2. MPH flow diagram.

*A. Single service request algorithm*

We start by considering a simple case with two sources ($K$=2), indexed by $j \in \{1,2\}$, both of which store a data item $l$ requested by destination $i$. If the size of the data item $l$ is $D_l$ and the maximum download time is $\tau_{il}$, then the total amount of bandwidth to allocate to the two VPNs (1→$i$ and 2→$i$) is $\widetilde{C}_{il} = D_l / \tau_{il}$. For the sake of simplicity, we denote $\widetilde{C}_{il}$ with $\widetilde{C}$ within a cycle. Clearly, a solution exists if $\bar{c}_{1i} + \bar{c}_{2i} \geq \widetilde{C}$, $C_i^T \geq \widetilde{C}$, and $C_1^M + C_2^M \geq \widetilde{C}$.

As regards VPN ($j$→$i$), the maximum amount of bandwidth which can be allocated is equal to $\bar{c}_{ji}^{\max} = \min(\bar{c}_{ji}, C_j^M, C_i^T)$. Let $c_{ji}^*$ be the amount of bandwidth already allocated to VPN ($j$→$i$); at the beginning of the first cycle $c_{ji}^*$ =0 for each theatre $i$.

The optimisation problem in (9) for each cycle can be very easily reduced to

$$F_{12}^i(\widetilde{C}) = \min_{\max(0,\widetilde{C}-\bar{c}_{2i}^{\max}) \leq c_1 \leq \min(\bar{c}_{1i}^{\max},\widetilde{C})}^+ [f_{1i}(c_1 + c_{1i}^*) + f_{2i}(\widetilde{C} - c_1 + c_{2i}^*)]. \quad (11)$$

The structure of equation (11) has appeared very frequently in literature under the framework of min-plus algebra; it is called the min-plus convolution [6]. The computation of (11) enables us to obtain $F_{12}^i$ (i.e., the minimum cost), which corresponds to a specific $c_1$ value that will be referred as $c_{1opt}$. Thus, the bandwidth to reserve is: $c_{1il}=c_{1opt}$ and $c_{2il} = \widetilde{C} - c_{1opt} = c_{2opt}$. Note that the min-plus convolution enables the optimum to be found for the specific request analysed. In fact, such an operation explores all possible admissible solutions and selects the best one. In the case of different solutions which provide an identical minimum cost, we force the minimum operator to first saturate a VPN (up to $\widetilde{c}_{ji}^{\max}$), and then to allocate the residual bandwidth ($\widetilde{C} - \overline{c}_{ji}^{\max}$) in the other VPNs. For this reason, we have denoted this modified minimum operator with $\min^+$. Thus, there are fewer mirrors with more available bandwidth, rather than more mirrors with fewer available resources to serve the subsequent service requests, and this may limit the number of VPNs to set-up.

Let us now increase the complexity and consider $K=3$. This case will require *two iterations*. The constraints become: $\sum_{j \in \{1,2,3\}} \overline{c}_{ji} \geq \widetilde{C}$, $C_i^T \geq \widetilde{C}$, and $\sum_{j \in \{1,2,3\}} C_j^M \geq \widetilde{C}$. If we assume we know $c^*_{1i}$ and $c^*_{2i}$, we can write the partial cost function $F_{12}^i(c)$ in the first iteration, where $c \in [c^*_{1i} + c^*_{2i}, c^*_{1i} + c^*_{2i} + \min(\widetilde{C}, \overline{c}_{1i}^{\max} + \overline{c}_{2i}^{\max})]$. In the second iteration, we proceed by computing

$$F_{123}^i(\widetilde{C}) = \min_{\max(0, \widetilde{C}-(\overline{c}_{1i}^{\max}+\overline{c}_{2i}^{\max})) \leq c_3 \leq \min(\overline{c}_{3i}^{\max},\widetilde{C})}^+ [f_{3i}(c_3 + c^*_{3i}) + F_{12}^i(\widetilde{C} - c_3 + c^*_{2i} + c^*_{1i})] \quad (12)$$

From (12), we obtain $F_{123}^i$ (the minimum cost), corresponding to a specific $c_3$ value that will be referred to as $c_{3opt}$. The values of bandwidth to allocate to the VPN ($3 \rightarrow 1$) is $c_{3il} = c_{3opt}$ and the total bandwidth to allocate VPNs ($2 \rightarrow 1$) and ($1 \rightarrow 1$) is $C_{12opt} = \widetilde{C} - c_{3opt}$. Finally, from $F_{12}^i(C_{12opt})$, it is easy to find $c_{1il} = c_{1opt}$ and $c_{2il} = C_{12opt} - c_{1opt} = c_{2opt}$.

The approach can be easily extended to the general case of $K$ sources. In this case, the minimum cost $F_{12...K}^i(\widetilde{C})$ corresponding to $c_{Kil}$ has to be computed, and then, on its return, all the other capacities, $c_{(K-1)il}...c_{1il}$, have to be allocated.

At the end of the optimal routing of a service request, the values $\overline{c}_{ji}$, $c^*_{ji}$, $C^T_i$, and $C^M_j$, which will be the inputs of the next cycle of the algorithm, have to be updated (see the step "Network status updated" in Fig. 2).

We stress that, looking at (11) and (12), it is clear that *for the case of the single request, this approach based on the min-plus algebra provides an effective way to implement an exhaustive search within the set of admissible solutions*.

*1) Complexity analysis for the single service request*

In this section we evaluate the computational cost of our heuristic algorithm. It is based on a number of cycles equal to $Q = \sum_l \sum_i u_{il}$, corresponding to the number of requests. Each cycle requires the computation of a number of min-plus convolutions which is bounded by $\left[\sum_j v_{jl} - 2\right]Z + 1$, where $Z$ is the maximum number of values, spaced by the unit interval $\Delta c$, used to represent the variable parameter of the min-plus convolution (e.g., $c_3$ in (12)), thus $Z = \max_{(i,l)}\{D_l / \tau_{il}\} / \Delta c$.

Clearly, the lower $\Delta c$ is, the higher the computational time.

Note that all the iterations, with the exception of the last, need to evaluate the min-plus convolution for all the admissible values of the independent variable $c$. Instead, the last iteration requires only one min-plus convolution for $c = \widetilde{C}$ to be computed.

To sum up, considering that the maximum number of steps needed to compute a min plus-convolution (as defined in (11)) is $Z$, the number of operations to compute $F_{12...K}^i(\widetilde{C})$ and the resource allocation in an MPH cycle is

$$\left\{\left[\sum_j v_{jl} - 2\right]Z + 1\right\} \cdot Z \leq K \cdot Z^2 \quad (13)$$

Since the MPH consists of $Q$ cycles, the complexity of computing all the min-plus convolutions is $O(QKZ^2)$.

In addition, the complexity to update the network status is $O(Q(K+N+KN))=O(QKN)$. At this stage, we indicate with $S$ the number of operations to perform the request scheduling (see Section III.B for details). Thus, the MPH complexity is

$$O_{MPH}=O(QKZ^2 + QKN+S). \quad (14)$$

In Theorem 3.1.6 (Properties of min-plus convolution for concave/convex functions) of [6], the Authors show that if $f_1$ and $f_2$ are concave with $f_1(0)=f_2(0)=0$, then the min-plus convolution $F_{12}(c)=\min\{f_1(c), f_2(c)\}$. Note that, a non-decreasing, concave function, null in zero, is also sub-additive, and thus non-decreasing, concave functions passing through zero are good candidates to represent the operational cost of a VPN.

In addition, since the min-plus convolution is associative and distributive over the minimum operator [17], it is easy to show that, given $K$ functions concave and passing through the origin, their min-plus convolution is:

$$F_{12...K}(c)=\min\{f_1(c), f_2(c), \ldots, f_K(c)\}. \quad (15)$$

Thus, it is possible to calculate $F_{12...K}(c)$ faster than by using the classic, direct computation of min-plus convolutions. The complexity reduces from $O(KZ^2)$ to $O(KZ)$. Please note that (14) can be applied only when the condition $\widetilde{C} \leq \min_j(\overline{c}_{ji}^{\max})$ holds. However, since such a condition may occur quite frequently, the computation of the bandwidth to be allocated would reduce to one step (i.e., $K$ operations). Thus, the MPH complexity can be reduced from $O(QKZ^2 + QKN+S)$ to $O(QK^3 + QKN+S)$.

*B. Service request scheduling*

In a homogeneous scenario (as defined in Section III), request scheduling does not influence the final solution. On the other hand, the schedule in heterogeneous scenarios for serving service requests by using the MPH algorithm can influence the final solution, and thus the cost $F_{TOT}$. Below, we introduce different scheduling algorithms, whose effectiveness in minimizing the global cost is analysed in Section IV. They are based on different sorting criteria, which are part of the scheduling algorithms themselves.

We first present a set of scheduling criteria based on the service demand only, characterised by the data item size and download time. In this case, a scheduling algorithm may be implemented as an initial sorting of requests according to the desired criteria; subsequently, the requests are selected according to their position in the order. As regards complexity, it is well known that the achievable complexity of a general sorting algorithm running over $y$ values is $O(y\log_2 y)$ [19]. Thus, $O(S)=O(Q \cdot \log_2 Q)$. These sorting criteria are listed below.

$D_d$ – service requests are sorted according to the data item size ($u_{il} \cdot D_l$) in descending order; in this case,

$D_a$ – service requests are sorted according to the data item size ($u_{il} \cdot D_l$) in ascending order.

$\widetilde{C}_d$ – service requests are sorted according to the amount of requested bandwidth $\widetilde{C}_{il} = u_{il} \cdot D_l / \tau_{il}$, in descending order.

$\widetilde{C}_a$ – service requests are sorted according to the amount of requested bandwidth, in ascending order.

*Rand* – service requests are randomly ordered; note that in this case, $O(S)=O(Q)$.

We then present another set of scheduling criteria, which depend not only on the service demand, but also on the data item catalogues in source nodes and their current available bandwidths. In this case, since the network status is updated at the end of each cycle, a single, initial sorting is useless. Thus, the selection parameter associated with each request must be computed each cycle and the best one is selected. As regards complexity, it is well known that the achievable complexity of a general selection algorithm running over $y$ values is $O(y)$. Thus, $O(S)=O(Q[Q \cdot K+Q])=O(Q^2 K)$. The selection criteria are listed below.

$N_d$ – the service request is selected according to the maximum number of source nodes (equal to $\sum_j v_{jl} \cdot u_{-1}(C_j^M)$) which store the requested data item and still have available bandwidth for allocation; $u_{-1}(c)$ denotes the unitary step function, equal to 1 when $c>0$ and 0 when $c \leq 0$.

$N_a$ – the service request is selected according to the minimum number of source nodes which store the requested data item and still have available bandwidth for allocation.

$C_d$ – the service request is selected according to the maximum total access bandwidth of the source nodes which store the requested data item, (equal to $\sum_j v_{jl} \cdot C_j^M$).

$C_a$ – the service request is selected according to the minimum total access bandwidth of the source nodes which store the requested data item, in ascending order.

$\hat{C}_d$ – the service request is selected according to the maximum, total bandwidth of the source nodes which store the requested data item (equal to $\sum_j v_{jl} \cdot C_j^M$) normalised by the bandwidth associated with the specific request, $\widetilde{C}_{il}$.

$\hat{C}_a$ – the service request is selected according to the minimum, total bandwidth of the source nodes which store the requested data item normalised by $\widetilde{C}_{il}$.

## IV. NUMERICAL RESULTS

In this Section, we evaluate the effectiveness of MPH in a homogeneous scenario, and then we extend the performance evaluation to heterogeneous cases.

The considerations made at the end of sub-section II.D about the characteristics of the objective function suggest that convergence to the global optimum cannot be guaranteed with standard solver for non-linear, convex problems, and that the usage of heuristics methods is justified. We have verified this conjecture by both analysing a simple, homogeneous scenario and solving the problem with the above mentioned commercial solver LINGO and MINOS. In the considered scenario, destination nodes access links, source nodes access links, maximum VPN capacities, data item sizes, download time requirements, source node data item catalogues and cost functions per-VPN are identical, and each destination requests one data item. In more details, this scenario models delivery of Digital Cinema movies [2] from $K$ mirrors to $N$ theatres, and is characterised by the following parameters: $D_l$= 200 GB with $l=1,…,H=16$, $C_j^M$= 1 Gb/s with $j=1,..,K$ ($K$=2, 3, 4 for the three different configurations analysed), $C_i^T$= 150 Mb/s with $i=1,..,N$=20, $\bar{c}_{ji}$ =150 Mb/s $\forall$ ($j,i$). We remark that the maximum VPN capacities are set equal to the destination node access speed, i.e., the core network is not the bottleneck. Each destination randomly performs a single request from the set of data item, and each source node has all data items in its catalogue. The download time is $\tau_{il}=\tau$ ($\tau$=3, 4, 5, 6 hours for the four different cases analysed).

With reference to (10), the VPN cost functions are homogeneous with $a_{ji}=a=1$ and $g_{ji}(c_{ji})=b_{ji} \cdot c_{ji}$, with $b_{ji}=b=0.01$ (Mb/s)$^{-1}$ $\forall$ ($j,i$). In this case, minimising the overall cost is equivalent to minimising the number of VPNs to set up.

Table I and Table II report the number of VPNs and the total cost obtained by LINGO and MINOS, with the maximum download time, $\tau$, and the number of source nodes, $K$, as parameters. Tables also report the global optima, which can be very easily computed by hand in this particularly simple case study, used to test the effectiveness of the tools. As expected, also in this trivial case study, LINGO and MINOS provide a higher number of VPNs (and thus a higher cost) than the global optimum for all configurations, with the exception of two cases (($K$=3, $\tau$=3h), ($K$=3, $\tau$=4h)) for LINGO and three cases (($K$=3, $\tau$=3h), ($K$=3, $\tau$=4h), ($K$=4, $\tau$=4h)) for MINOS. When $K$=2, due to the constraint (1), it is impossible to respect the maximum download time set at 3 and 4 hours, and thus the problem cannot be solved.

As for MPH, the result is that it can always find the optimum values of the VPN number and total cost reported in Tables 2 and 3, respectively. This is an expected result in homogeneous scenarios, since the MPH can find the optimal solution (with the minimum number of VPNs) for the single service request; in addition, it can also perform a resource allocation (via the min$^+$ operator) at each step, so as to concentrate the access bandwidth availability in a minimal subset of source nodes, thus allowing the minimisation of the number of VPNs for serving future requests. Finally, in homogeneous scenarios, we remark once again that the scheduling scheme does not have any impact on the solution.

In a further experiment, we enabled the LINGO option for using the GLOBAL OPTIMUM (GO) search criterion, instead of using default configuration suitable for convex optimisation. With this setting, the solver proved to be able to reach the global optimum, at the expenses of a strongly increased computation time.

TABLE I. NUMBER OF VPNS (LINGO VS. MINOS VS. GLOBAL OPTIMUM)

| Download time, $\tau$ (hours) | K=2 | | | K=3 | | | K=4 | | |
|---|---|---|---|---|---|---|---|---|---|
| | LINGO | AMPL/ MINOS | Global optimum | LINGO | AMPL/ MINOS | Global optimum | LINGO | AMPL/ MINOS | Global optimum |
| 3 | - | - | - | 22 | 22 | **22** | 22 | 22 | **20** |
| 4 | - | - | - | 20 | 20 | **20** | 22 | 20 | **20** |
| 5 | 21 | 21 | **20** | 21 | 21 | **20** | 21 | 21 | **20** |
| 6 | 21 | 21 | **20** | 21 | 21 | **20** | 21 | 21 | **20** |

TABLE II. TOTAL COST (LINGO VS. MINOS VS. GLOBAL OPTIMUM)

| Download time, $\tau$ (hours) | K=2 | | | K=3 | | | K=4 | | |
|---|---|---|---|---|---|---|---|---|---|
| | LINGO | AMPL/ MINOS | Global optimum | LINGO | AMPL/ MINOS | Global optimum | LINGO | AMPL/ MINOS | Global optimum |
| 3 | - | - | - | 51.6296 | 51.6296 | **51.6296** | 51.6296 | 51.6296 | **49.6296** |
| 4 | - | - | - | 42.2222 | 42.2222 | **42.2222** | 44.2222 | 42.2222 | **42.2222** |
| 5 | 38.7778 | 38.7778 | **37.7778** | 38.7778 | 38.7778 | **37.7778** | 38.7778 | 38.7778 | **37.7778** |
| 6 | 35.8148 | 35.8148 | **34.8148** | 35.8148 | 35.8148 | **34.8148** | 35.8148 | 35.8148 | **34.8148** |

In more detail, the average solution time for default configuration results in the order of few seconds, whereas enabling the GO flag increases it up to hundreds of seconds for this trivial and size-limited configuration.

Any further increase of the system size makes the solver no longer effective. In fact, if the number of both source and destination nodes is scaled by a factor equal to 10, the solver requires about 45 minutes to provide the first admissible solution. Note that we are not referring to the global optimum, since the solution found has a cost larger than that found by using MPH. For any further increase of the system size, the solver, after being running for 3 hours, has not been able to provide even an admissible solution.

Instead, the computation times provided by MPH remains in the order of few seconds, setting the unit interval parameter $\Delta c$ equal to 1 Mb/s. These results definitely confirm the inadequacy of LINGO with this option to solve real scale problems. As for MINOS, it does not provide a similar feature for global optimizaton. In the following, we will compare MPH with LINGO without the GO criterion enabled, since this is the only configuration which makes the solver usable for this problem also on larger system sizes with affordable computation times.

In the next case studies, the *network* scenario is still the one described above. There are $K$=4 source nodes, with an access capacity of 1 Gb/s, $N$=20 destination nodes, with an access capacity of 150 Mb/s, and a maximum VPN capacity of 150 Mb/s. Heterogeneity is put on the film catalogue, service demand, and VPN cost functions.

As regards the data item catalogue, the number of data item is equal to $H$=40, the size of each is modelled as a random variable uniformly distributed in the range from 130 GB to 270 GB (values compliant with Digital Cinema scenario); each data item is present in 3 source nodes, thus each mirror stores 30 data item (approximately 4 TB of data per source nodes). The data item catalogue has been assumed to be identical in all the experiments we performed.

As regards the service demand, each theatre requires $H_R$=2 data item each time, randomly selected from the data item catalogue, with a download time, $\tau_i$, for both data items. $\tau_i$ is modelled as a random variable uniformly distributed in the set of integers ranging from $\tau_{min}$=6 hours to $\tau_{max}$=12 hours. In order to do an exhaustive analysis of the performance of the solving algorithms, it would be necessary to evaluate the cost values and number of VPNs averaged over all the configurations of the service demand. Since exploring all configurations is infeasible (it is easy to verify that they would be equal to 8575[25]), we have run MPH over a sample of $10^4$ configurations, and computed the average of the total cost values and of the number of VPNs for all the scheduling algorithms presented in the previous section, as well as for the Lingo solver (labelled as LS). The LS performance proved to be very close to that of MINOS, which is not reported to improve figure neatness.

For the considered configuration, it is possible to give a rough estimation of the computing power needed to perform an exhaustive search on the whole problem (the second option sketched at the beginning of section III). Assuming to use for the content size the average value $\overline{D}$=200 GB, and for the download time the average value $\overline{\tau}$=9 hours, a brute force analysis considering all requests simultaneously means exploring a number of configurations $N_c$ equal to:

$$N_c = \binom{K \times C^M / \Delta c}{\lceil \overline{D}/(\Delta c \times \overline{\tau}) \rceil \times N \times H_R} = \binom{4000}{2000}, \quad (16)$$

which is definitely not feasible.

As regards the VPN cost functions, we assume them to be in the form of (10) with $g(c)=b \cdot c$. We have analysed three different scenarios:

*Scenario 1*: all VPN cost functions are equal, with parameters $a$=1 and $b$=0.01 (Mb/s)$^{-1}$;

*Scenario 2*: all VPN cost functions are equal (with parameters $a$=1 and $b$=0.01 (Mb/s)$^{-1}$), with the exception of those connecting destination nodes to source node 1, with $a$=3 and $b$=0.03 (Mb/s)$^{-1}$;

*Scenario 3*: destination nodes and source nodes are partitioned into two subsets of equal cardinality, named $T_1$ and $T_2$ for destination nodes and $M_1$ and $M_2$ for source nodes, respectively. The VPNs which connect $T_1$ to $M_1$ and $T_2$ to $M_2$ have a cost function characterised by $a=1$ and $b=0.01$ $(Mb/s)^{-1}$, whereas those connecting $T_1$ to $M_2$ and $T_2$ to $M_1$ have $a=3$ and $b=0.03$ $(Mb/s)^{-1}$.

In scenario 1, there is not any preference on the choice of source node(s) for all destination nodes. The scenario 2 may represents a distribution system with a source node accessible through an expensive connection provided by the SON, to be used only when the other source nodes are overloaded. Finally, the scenario 3 models a distribution system with a preferential pre-association (e.g, on the basis of geographical or IP distance) between groups of destination nodes with groups of source nodes.

The maximum processing time for the network configurations analysed is in the order of few seconds for LINGO, and few milliseconds for the heuristic approach (implemented in C++) on an standard PC (Intel Core2 Duo @2.2 GHz equipped with 2 GB of RAM).

Fig. 3 is relevant to the scenario 1 and illustrate the estimation of the cumulative probability distribution function (CDF) of the total cost function, obtained from the $10^4$ configurations. The number of samples equal to $10^4$ appears sufficient to have a good estimation of the CDF, since the coefficient of variability (standard deviation divided by the average value) is always lower than 6%. Fig. 3 shows that LS performance is worse than MPH, whichever scheduling algorithm is used. Similar performance is observed for the scenarios 2 and 3, not shown due to space limitations.

In order to investigate which are the best scheduling algorithms for the three scenarios, Fig. 4 shows the percentage of runs in which MPH outperforms LS in terms of total cost, for all the scheduling criteria and for all scenarios. It appears that the best ones are those depending not only on the service demand, but also on the data item catalogues in source nodes and current available bandwidth. In general, the schemes $N_d$ and $N_a$ (based on the number of source nodes storing the requested data item) provide a total cost lower than that of LS in more than 94% of runs in all scenarios; $N_a$ shows the best performance (approximately 96% on average), whereas $N_d$ performs slightly worse (approximately 95% on average). As for the scenario 3, the best scheduling algorithm is $C_d$ (based on the maximum total access bandwidth of the source nodes which store the requested data item), which provides a cost lower than LS in 96% of runs.

Fig. 5 gives quantitative details concerning the gain of MPH over LS in terms of cost (Fig. 5.a) and number of VPNs (Fig. 5.b). Also for these performance figures, the three best performing scheduling approaches are $N_d$, $N_a$, and $C_d$. For instance, in the scenario 2, the gain of $N_a$ is next to 12% for the cost and 20% for the number of VPNs.

As regards the solving capabilities of LINGO, we also remark that five runs out of 30,000 did not produce any admissible solutions after several hours of processing time, whereas MPH always succeeded in a few milliseconds.

We can conclude that, on average, MPH outperforms LINGO independently of the specific scheduling scheme. Moreover, the scheduling algorithms which updates the selection parameter in each MPH cycle are the best performing ones.

This behaviour has also been confirmed in other system configurations (network scenario, service demand, cost functions, and film catalogues on mirrors); the relevant quantitative results are not reported in this paper due to space limitations.

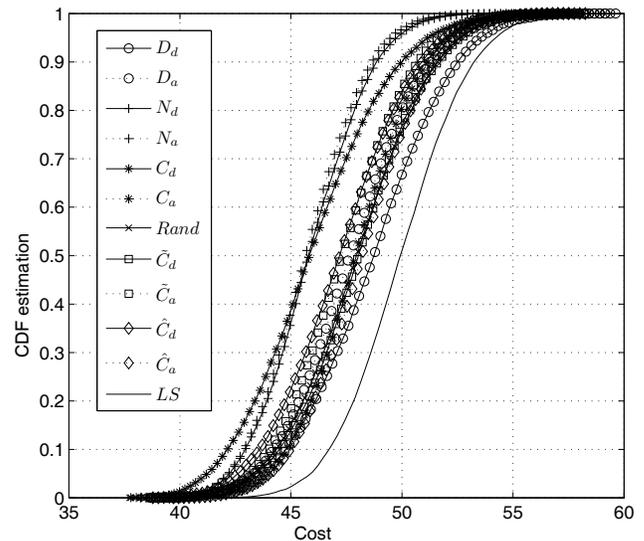

Fig. 3. Estimation of the CDF of network cost for all scheduling algorithms in the scenario 1

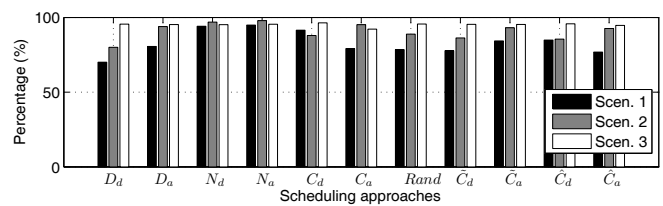

Fig. 4. Performance analysis of MPH for each scheduling algorithm: percentage of runs performing better than Lingo in terms of cost.

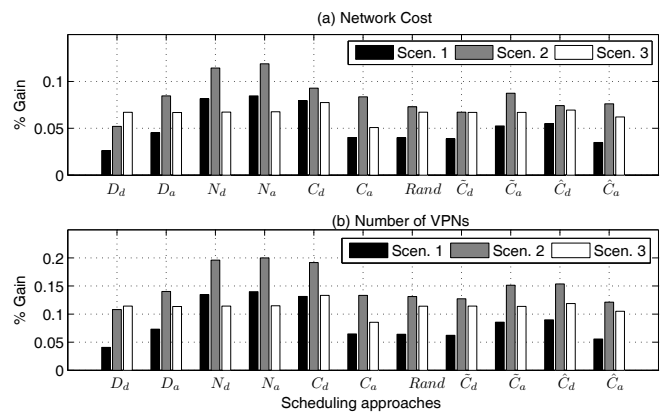

Fig. 5. Average performance improvement of MPH for each scheduling algorithms with respect to Lingo, in terms of both cost (a) and number of VPNs (b).

In addition, in the specific scenarios that we have considered, the MPH with the $N_d$, $N_a$, and $C_d$ scheduling schemes prove to be the best ones. Nevertheless, this may not be true in a general configuration. Even if these results do not hold for all possible application scenarios, many practical situations that can be found in operation can be led back to one of the analysed scenarios. In any case, since the time needed to execute the MPH algorithm is definitely low (in the order of a few milliseconds for this simple type of case study, up to few seconds for larger configurations), depending on the timing of the considered service environment, a practical way to proceed is to solve the problem by using different scheduling schemes and then to pick up the solution which provides the lower cost.

One could argue that that in a slightly more realistic setting, service requests will come and go and resources are allocated and released accordingly (on-line problem), whereas in our approach they are packed and then resources are allocated to all requests together (off-line problem). We note that, if the arrival times are close to each other so that there are no requests that have completed the download at the arrival of the last request, solving this problem in a first-in first-out order corresponds to adopt a random selection strategy (*Rand* in section III.B) for what concerns the value of the overall cost function. Since the *Rand* strategy is not one of the best performing ones, whenever it is possible formulating the problem as an off-line, it is convenient doing so.

V. RELATED WORK

There are a number of related works in the field, which we can roughly classify in server selection problems, planning solutions, non guaranteed delivery solutions, peer-to-peer (P2P)-based solutions, and multicast-based solutions.

We denote as server selection problems those treated in works such as [30], where the objective is to select the most suitable server to serve each client. However, in these schemes, usually the parallel downloading defined in [5] is not employed. Thus, these proposals are definitely less effective than ours. In addition, most of these schemes are not able to provide guaranteed delivery times. In this regard, it is worth to mention also the recent paper [27], which illustrates an innovative strategy to assign clients to servers. Nevertheless, as in other proposals of this kind, this solution cannot provide guaranteed delivery times. Thus, we can see our solution as an enhancements of these scheme, able to benefit of both guaranteed resources and parallel downloading capabilities.

We denote planning solutions those dealing with the provisioning of either bandwidth and storage resources in the SON nodes in order to support the distribution of bulk data service over the Internet, such as [20]. Clearly, in this case the perspective is completely different, since the goal is to design an overlay network (provisioned with bandwidth and storage resources) able to efficiently use resources to effectively support bulk data transfer service.

NetStitcher [26] is an excellent example of a solution designed to support bulk data transfer without guaranteed delivery. This solution usually benefit of unused resources and proposes a store-and-forward of data contents through intermediate nodes, to save backbone bandwidth at peak hours. However, this kind of solution, even if extremely efficient, is not able to a priori guarantee delivery times in arbitrary time windows, and thus it is not suitable for the analyzed problem.

One may wonder if a P2P architecture or multicast distribution may be valid alternatives to our model. Let us start evaluating the first option. In the considered scenarios, a completely distributed P2P architecture cannot represent a complete alternative to centralized-decision delivery, since it is not able to guarantee the desired QoS yet [4]. For this reason, we have introduced the role of the SON provider, since the content operator can guarantee bandwidth only on the access connections of his servers. Finally, using the SON also in a P2P setting to guarantee QoS would require a possible huge number of virtual paths, with relevant cost explosion.

All these considerations enforce the idea that in the considered scenario (guaranteed delivery), a centralized decision approach for content distribution is the best solution. On the other hand, it is also worth noting that in the last few years, the P2P research community has produced a large number of interesting results in content distribution. As highlighted in [3], it is very important to think about how centralized solutions and P2P solutions can coexist and collaborate to offer users the best performance. A future investigation should be carried out into how and whether a P2P system can represent an improvement in the proposed basic architecture. In particular, it is evident that, already in the present form, this work can be employed in a P2P system with centralized decisions. In fact, let us consider not an whole data item, but data item chunks, and define a delivery phase as the time needed to complete a data item chunk download. Thus, after the completion of a delivery phase, the number of source nodes is increased, and the MPH algorithm can be re-run, taking into account also new potential sources, so building a system similar to that described in [25], but able to provide performance guarantees and cost minimization. In fact, please note that this P2P-oriented usage of MPH would always take into account already set-up VPNs to minimize network cost.

Finally, as for the usage of IP multicast, it seems to be not a good choice due to the following reasons:

- multicast protocols are not widespread, especially with the requested reliability features for the guaranteed delivery of bulk data. Although, the IETF has done some steps towards the definition of a standard framework for provider-provisioned multicast VPNs [18], widespread deployment is still far from reality;
- the access link bandwidth of destination nodes may differ (heterogeneous scenarios). This means that the slowest access connection would determine the overall transfer time, or multiple multicast groups should be used, collecting "homogeneous" destination nodes, and thus strongly mitigating multicast benefits.

A more viable option for multicast-based solutions is to implement them in SON nodes only. Thus, possible candidates for supporting new generation of guaranteed delivery of bulk data could be application layer multicast solutions [31][32]. Another slightly different alternative to provide the needed functions (guaranteed bandwidth, reliability) in an overlay network with multicast capabilities could be realized using programmable network devices, such as those supporting the OpenFlow protocol [33], coupled with platforms supporting network services with packet processing capabilities, such as those provided by the NetServ [34], two features allowing to easily introduce innovation in the network.

VI. CONCLUSION

In this paper we have analysed the problem of bulk data transfer with guaranteed delivery times. We have considered the perspective of a generic content/service operator. We have assumed that content is present in more than one source node and that the operator has to simultaneously manage a number of bulk transfers, with some source nodes contributing to more than one download. We have modelled this distribution system as a continuous and non-linear optimization problem. We have defined an heuristic named MPH, based on the min-plus algebra, able solve the problem in acceptable times and with high effectiveness.

Numerical results show that the proposed heuristic outperforms two well-known commercial solvers we tested on a number of sample configurations, whilst maintaining an affordable computational burden, which is indeed definitely lower than that of the commercial solvers.

Finally, we have discussed how the proposed solution can be also used in controlled P2P systems, and/or new application layer multicast architectures.

Future works will extend the work providing the calculation of a lower bound to the optimal cost value, usable to further reduce computation times by introducing a stop condition, and further investigating on more refined request scheduling algorithms.

In addition, we will develop a system prototype. The VPN endpoints will be implemented by using programmable nodes operating according to the NetServ architecture [34]. The central signalling node, which is in charge of collecting service requests and executing the MPH algorithm, will be implemented through a service running on the Mobicents JSLEE server (JAIN service logic execution environment), a carrier-grade telecom software platform [35][36]. This platform is particularly suitable for this task, since it has been recently equipped with tools able to ease the design of complex signalling services [37].